# How to evaluate games in education: a literature review

Giulio Barbero[1,*], Marcello M. Bonsangue[1], and Felienne F.J. Herman[2]

[1] LIACS, Leiden University, Leiden, The Netherlands
{g.barbero, m.m.bonsangue}@liacs.leidenuniv.nl
[2] Network Institute, Vrije Universiteit, Amsterdam, The Netherlands
f.f.j.hermans@vu.nl

**Abstract.** Adding game elements to higher education is an increasingly common practice. As a result, many recent empirical studies focus on studying the effectiveness of gamified or game-based educational experiences. The findings of these studies are very diverse, showing both positive and negative effects, and thus calling for comparative meta-studies. In this paper we review and analyze different studies, aiming to summarise and evaluate controlled experiments conducted within different scientific disciplines. We focus on the clarity of non-experimental conditions' descriptions and show that in most cases (a) educational methods used in control groups' activities are poorly described, (b) educational materials used in control groups' activities are often unclear, and (c) the starting conditions are unclear. We also noticed that studies in the fields of computer science and engineering, in general, report results more clearly than in other fields. Based on the above finding, we conclude with a few recommendations for the execution of future empirical studies of games in education for the sake of allowing a more structured comparison.

## 1    Introduction

The digitalization of education is a common trend in many different contexts. Just to cite a few examples, the amount of courses available online is countless, most books also make a digital version available, and a massive amount of lectures are streamed and available online every day. Digitalization opens the way to new methods to convey information and its applications in education creating spaces for new teaching tools and techniques. A very promising one is extracting elements from video games and their application to enhance education (*gamification*). Moreover, video games can also be used in their entirety to convey scholastic knowledge (*game-based learning*).

In this paper, we collect and analyze studies involving controlled experiments on gamified and game-based learning techniques applied to scientific secondary or higher education. When gamification is applied in this context, experimentation often translates as the application of playful elements in education. On the other hand, game-based learning usually involves the design of video games fully focusing on conveying educative content. The goals of these experimental tools can vary from enacting behavioral changes [1, 2] to improving knowledge acquisition. In fact, the analysis in [3] reports knowledge acquisition as the predominant learning outcome of 64 studies out of 105.

Regardless of the field of application, the cognitive effects of games or game elements vary as well. The most commonly reported effect is an increase in students' motivation and engagement, in line with the historical trend to utilize games in education as a medium to make experiences more pleasurable [4, 5]. On the other hand, the link between game elements and actual education effectiveness is not as clear. Even though many studies report a positive relationship between the two [6] and are backed by empirical game research [7], many experiments present opposite results [8] with theoretical research to support them [9].

Such diversity of studies (and results) in the field justifies the relatively high number of meta-studies. These tend to focus on specific aspects of the gamified/game-based learning experiments but also include more general information about the contexts of application (where possible). For example, the review in [3] focuses on empirical studies between 2000 and 2013 that apply game-based learning in primary education with the goal of documenting the diversity in the field in terms of learning outcomes, the topic of application, and the quality of the study. Other works focus on game motivators, how games are used in education, and their effect [10]. While the selection criteria of the latter are less strict compared to [3], it includes also more recent papers (from 2000 to 2019). Finally, [11] performs a broader review, focusing on how the studies were carried on. Through this perspective, they draw several conclusions about the quality of the experiments. In particular, they define multiple issues arising from the lack of clarity in the reporting style of many studies. They also report a sharp increase in studies involving gamification which more than quadrupled in one year (2011-2012).

The lack of clarity in the reporting style of many experiments makes comparative approaches challenging. In particular, the impact of different contexts of application related to the game components implemented. The relevance of the context can become especially evident in controlled experiments where the effect of experimental conditions (i.e. the game elements applied to education) can usually be better analyzed with a clear understanding of the control conditions (i.e. the standard education method for that specific context). However, this is often challenging or impossible due to a lack of clarity in the description of the control conditions [11]. In practice, the lack of this type of contextual information makes it difficult to evaluate the experimental approach; how is the subject taught in the control group? What type of material is used? How are the experimental and control groups evaluated?

## 2    Research Space Definition

Our research question is derived and influenced by two main characteristics of the field: diversity in the type of studies and diversity in the context of its application.

**Characteristic 1: Type of studies** Many empirical studies that utilize games involve the use of an experimental and a control group. Although information about control (or no-game) conditions can be helpful to evaluate findings, many studies in the field omit it to different degrees. This is a known issue that can hinder analytical approaches focused on the influence of experimental conditions [12]. Other studies do not use control groups and only rely on qualitative analyses of information gathered over the entire population. Categorizing these studies is even more complex and comparison with different studies is challenging. Based on this initial difference, we select papers based on the presence of a control group. Following the standard academic path, this criterion ensures the relevance of pre-intervention context descriptions (also when provided by the same course results in previous years).

We also determine common elements that are necessary to replicate a controlled experiment involving the use of gamification and game-based in education. We categorize each paper by reporting how much information we can find about starting conditions usually represented by the control group. In this regard we define three elements as relevant: type of teaching material (the tools used to transmit course content), teaching method (how is the course taught), and evaluation method (how is the effectiveness of experimental and control methods evaluated).

**Characteristic 2: Context of application** Games have been studied and used in education throughout different academic curricula, from scientific to humanistic subjects, in academic and technical education. In the study of languages, for example, game elements have been used and appreciated in both academic and "more commercial" settings [13, 14]. Also, the field of mathematics experimented with adding game components to different grades of education [15]. Moreover, games are used and studied for both practical (training) and theoretical knowledge acquisition. Finally, another large part of the studies involves the use of game elements in behavioral change projects aimed at educational environments, for example, to promote safer sexual practices [16]. Such diversity naturally arises from the shared interest in the use of educational games but makes comparisons between studies very challenging. Therefore, we narrow our search by using a few strict criteria in selecting the studies to review. First, we focus on studies about how games influence the absorption of scientific notions. Since we focus on how different fields produce different studies, we include both natural and social sciences in order to preserve some variation. Also, we include only studies that aim to the acquisition of theoretical or practical knowledge. The rationale behind this is to include studies in the field of medicine and nursing which often mix the two. On the other hand, we exclude studies whose goal is to develop behavioral changes. Finally, we focus on studies involving secondary education. This includes high school and university-level courses. It excludes doctoral and specialization studies in which participants are often professionals.

## 3   Method

We summarise and motivate here further criteria used to filter the studies. We have chosen to focus on studies presented in English and to continue from where [3] left off, considering works from 2013 to 2020. We decided to focus only on recent development in the field because of the increased number of game-based experiments in education [11]. We consider only video games and hybrid games, in light of the fact that the biggest majority of the studies collected involve digital components. In this way, we want to eliminate the few outliers which could prove difficult to compare. Of course, all experiments must involve the use of game elements or full games. This excludes pure simulations in which game systems, or more generally "pleasurable" components, are not implemented.

We implemented the above criteria by searching the Leiden University Library[1] using the query ("serious game" OR "game-based" OR "gamification" OR "game elements") AND ("experiment" OR "evaluation" OR "impacts" OR "outcomes" OR "effects" OR "education" OR "learning").

After using the above search we proceeded by reading the abstract of each of the first 100 papers sorted by relevance. We then determined whether it respects the rest of the aforementioned selection criteria (double-checking the year of publication). Studies not respecting one or more of the above criteria have been excluded from the review. At the end of this selection process, we collected 89 studies. We further read the full paper and made a final selection of 43 relevant studies.

With the goal of documenting diversity and clarity of starting conditions in mind, we summarise the categories through which the studies are classified:

- Field of application: medical sciences (medicine and nursing), natural sciences (biology, mathematics, physics, computer science, etc.), economics (economics, business, management), and social sciences (sociology, psychology, anthropology).
- Type of education: theoretical, practical (training), or a mixture of both.
- Clarity in the context of application: type of teaching material, teaching method, or evaluation method. For each, we assign the type "Unclear" if the experiment cannot be replicable with the information presented, and "Clear" otherwise.
- Grade of education: high school level or university level.
- Results: The effect of the experimental condition on motivation and performance. We consider "Positive", "Negative", and "Mixed/No-change" effects depending on the quality of the result in game condition.

## 4   Results

---

[1] https://www.library.universiteitleiden.nl/

In this section, we report the quantitative results derived from the analysis of the included studies through the aforementioned categories. Of the 43 studies we collected[2] almost half (N=20) are quite recent (i.e., published in 2017 and 2019) and 9 were published in 2020. The average publication year is 2018.

Medicine is the topic with the highest number of studies (N=8), followed by math (N=6) and computer science (N=6). All the other topics score equal to or smaller than 4. The number of studies which investigate the effect of game-based education also (at least in part) on training is 10, of which 8 are contextualized in courses involving a life science (medicine, nursing, or physiotherapy)(see Figure 1.1).

We cluster the fields reported in the studies into four categories: hard sciences, life sciences, social sciences, and engineering. For each category, we then calculate the success rate. The two categories with the highest success rate are hard sciences (N=10/18) and engineering (N=3/4), with computer science being the subject with the highest success rate (N=4/6). It is important to remark that there is some possible overlap between these two categories (for example, subjects studied in computer science could be also studied in computer engineering). Life sciences present the lowest success rate (N=5/14). Almost the majority of the studies reported "positive" results (N=21/43) while 12 studies reported "mixed" results.

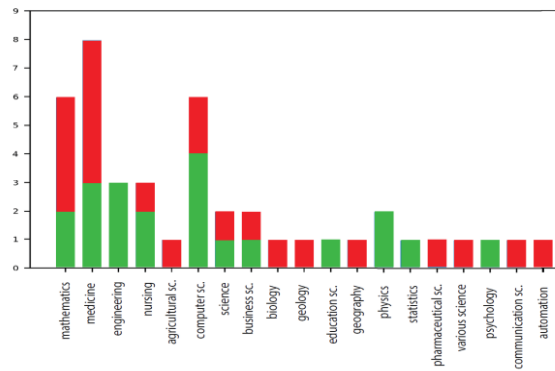

**Fig. 1.1** Number of studies per subject. The positive success rate in green, mixed and negative in red.

We analyzed the frequency of the scores for the details of the educational context. For 22 studies the details reporting the type of teaching material used in the control groups are deemed unclear (see Figure 1.2). In the case of the teaching method, the results show that in 30 studies this information is also unclear. On the other hand, testing methods are often more clearly reported with only 12 studies marked as unclear. Only 6 studies lack detailing in all three categories while 7 studies are clear in all three. Some are unclear in one or two categories (for both cases N=15).

---

[2] Including [17, 18, 19]. See the Appendix for completing the list of references to the studies considered.

Almost half of the studies (N=21) report positive results across both performance and student attitudes (when this was relevant). 12 studies report mixed results, noticing only partial improvement in either performance or student attitudes. The rest (N=10) report worse results in the conditions involving games.

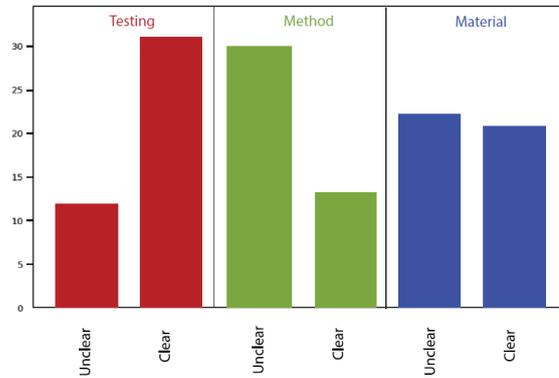

**Fig. 1.2** Number of studies per clarity score

The Pearson correlation coefficient between the success rate (mapping the scale 'negative' - 'mixed' - 'positive' over a scale from 0 to 2) and the average of the context clarity values (considering the 'clear' label as 1 and the 'unclear' as 0) for each reviewed article shows no meaningful correlation between the two (r=0.073). The correlation coefficient for the individual clarity of educational context values and the success scores also does not report a meaningful correlation (teaching material r= -0.078, teaching method r=0.229, testing method r=0.004). We additionally calculate the correlation coefficient between the values for the clarity of the educational context. Also in this case, no meaningful correlation is reported (material-teaching method r=0.369, material-testing r=0.089, testing-teaching method r=-0.042).

Within the studies we found lacking in terms of detailing scores, we delineate different groups based on how the lacking aspect is treated. This creates two categories valid for both the scores related to the clarity in reporting teaching material and method. One category includes studies that simply did not mention these two components [17, 18]. The other category does mention these elements of the control conditions but resolves them in generic terms (e.g.: defining them as "standard", "classical" or "traditional") [19, 20].

The same categories cannot be applied to the scores of the evaluation method detailing. The main reason is that these studies are more consistent. Even when being "unclear", the evaluation is mentioned in the paper as an essential part of the experimental design. However, these cases refer to their results (e.g.: the presence of an "improvement") without explaining in detail the type of tests and analyses that led to those conclusions.

## 5 Discussion

The present review focuses on how information in controlled studies involving games in education is presented. Our analysis shows that the most frequently unclear element of the educational context is the teaching method. In this regard, many studies fail to mention the way the teaching material is presented. It is important to notice that, when it comes to this teaching material, the actual information provided is more complete than the information regarding the teaching method; many studies report at least the type of material used (specific books, PowerPoint slides, and online courses are the most common).

Following also the qualitative parts of our analysis, we cluster those studies which are unclear according to these first two scoring systems and postulate reasons for the lack of detailing. Some of these studies rely on the idea that the respective educational practices (teaching method and material in this case) are supposedly standardized in certain contexts. However, this type of standardization often does not account for teaching methods and/or the variation of these courses throughout the years in terms of content and material used. Other studies, in particular regarding the educational method, might simply overlook the importance of these elements and do not touch the subject.

The description of the testing method is usually clearer and in line with expectations. Few studies score 0, usually focusing on additional qualitative evaluations after the actual performance tests. Overall, it is rarer to encounter studies reporting no information at all in this regard, probably due to the intrinsic and fundamental nature of evaluation methods in experimental designs.

Looking at the success rate, results show that many studies (N = 21) reported benefits from using games for parts of their educational components. However, looking at the success score for each topic, games or game elements do not seem to be equally effective in every subject; for example, only three studies in the field of medicine report "positive" results (N = 3/8). This could indicate that game elements are more effective in some fields compared to others. Topics that often heavily involve the study of technology (engineering and computer science) score instead above average (respectively, N = 3/3 for engineering and N = 4/6 for computer science). This could indicate that students that are usually more involved, or at least interested, in technology are more susceptible to the effects of games or more motivated by such tools. However, other subjects that make heavy use of computational tools, such as mathematics, show a low success rate (N=2/6).

Finally, it is important to observe that the lack of correlation between the scores (in particular detailing scores and success scores) can in part attest to the research's quality since the final results of the experiment do not seem to influence reporting decisions.

**Limitations** Our inclusion criteria are effective in defining the area of interest of our research. However, they also present some intrinsic limitations. For example, our analysis focuses on subjects related to natural, social, and technical sciences. This is functional to collect studies that follow a common scientific experimental method, but it excludes a relevant amount of studies, in particular in the fields of language acquisition and history education. Also, by selecting science we automatically excluded studies in purely professional training settings. This often (but not always) affects experimental applications to training courses in vocational schools. Another limitation comes from the subjectiveness of the clarity scoring system we used, based on our own interpretation of the various texts. Finally, the success scoring system (negative-mixed-positive results) is highly dependable on the scope of the individual study. Some focus and report exclusively on performance improvement, others might include students' engagement and motivation. However, it is a valid parameter to determine how those results are analyzed in each study and what perspective each paper takes on game-based education effectiveness.

# 6     Conclusions

This paper highlights the lack of clarity in describing the educational context as a common issue in the field of game-based education studies. This has an impact on the difficulty to interpret exactly what educational elements the gamified/game-based experience replace. Subsequently, it strongly hinders comparative analyses. Moreover, studies with incomplete information make it very challenging to perform any in-depth categorization according to the educational context. We also hint at a relationship between the success of gamified/game-based education and students' intrinsic motivation.

Basing ourselves also on the aforementioned limitations, there are several directions to extend this work. One could use the same method and apply it to studies in the field of humanities. This is a better option than incorporating these studies directly in the current selection since the two groups present very different teaching goals and methods. It is also relevant to apply the same method to studies in vocational education using clarity scores more closely linked to practical education. Finally, comparative studies should take into account the diversity in the completeness of information, developing a methodology that can either adapt to it or compare different perspectives.

# Appendix

In this appendix, we list all papers collected and analyzed in our review that have not been included in the references.